
Apparently-To: <HEP-TH@XXX.lanl.gov>

\input harvmac
%
%
\def\today{\ifcase\month\or
   January\or February\or March\or April\or May\or June\or
   July\or August\or September\or October\or November\or December\fi
   \space\number\day, \number\year}
%
%
%
%
%
\def\dooley{\centerline{Hyuk-jae Lee\footnote{$^*$}
{Bitnet: lhj5551@krysucc1}}\bigskip
\centerline{{\it Department of Physics, Yonsei University}}
\centerline{{\it Seoul 120-749, Korea}}\bigskip}
\def\submit{\baselineskip=20pt plus 2pt minus 2pt} 
           \def\c{\chi}       
     \def\f{\phi}       
  \def\g{\gamma}           
\def\l{\lambda}   \def\L{\Lambda}             
\def\r{\rho}          \def\o{\omega}     \def\O{\Omega}
\def\p{\psi}      \def\P{\Psi}        \def\s{\sigma}     
            
               \def\w{\wedge}
        
%

\def\CA{{\cal A}}   
\def\CD{{\cal D}} \def\CF{{\cal F}}  \def\CG{{\cal G}}
\def\CH{{\cal H}}   
\def\CO{{\cal O}}   \def\CQ{{\cal Q}}
   
\def\CW{{\cal W}} \def\CM{{\cal M}}  \def\CS{{\cal S}}
 \def\CO{{\cal O}}
%
\def\rd{\partial}

\def\darr#1{\raise1.5ex\hbox{$\leftrightarrow$}\mkern-16.5mu #1}

\def\Fr#1#2{{#1\over#2}}

%
\def\cmp#1#2#3{, Comm.\ Math.\ Phys.\ {{\bf #1}} {(#2)} {#3}}
\def\pl#1#2#3{, Phys.\ Lett.\ {{\bf #1}} {(#2)} {#3}}

\lref\wit{E. Witten\cmp{117}{1988}{353}}
\lref\lab{J. M. F. Labastida and M. Pernici\pl{B212}{1988}{56}}
\lref\aj{M. F. Atiyah and L. Jeffrey, J. Geom. Phys. {\bf 7}(1990)120}
\lref\singer{L. Baulieu and I. M. Singer, Nucl. Phys. (Proc. Suppl)
            {\bf 5B}(1988)12}
\lref\ken{H. Kanno, Z. Phys.{\bf C43}(1989)43}
\lref\ati{M. F. Atiyah and I. M. Singer, Proc. Natl. Acad. Sci. USA
{\bf 81}(1984)2597}
\lref\itoi{M. Itoh, Proc. Japan Acad. {\bf 59}(1983)431}
\lref\itoii{M. Itoh, Osaka J. Math.{\bf 22}(1985)845}
\lref\kim{H. J. Kim, Math. Z. {\bf 195}(1987)143}
\lref\park{J. S. Park, {\it $N=2$ Topological Yang-Mills Theory on
Compact K\"ahler Surfaces},Preprint}
\lref\bir{D. Birmingham, M. Blau, M. Rakowski and G, Thompson, Phys. Rep.
{\bf 209},(1991)129}
\lref\gal{A. Galperin and O. Ogievetsky\cmp{139}{1991}{377}}
\lref\witt{E. Witten, Preprint IASSNS-91/83}
\lref\bra{S. B. Bradlow\cmp{135}{1990}{1}}

\Title{\vbox{\baselineskip12pt\hbox{YUMS 93-16}\hbox{SNUTP 93-45}}}
{\vbox{\centerline{Topological Field Theories on}
    \vskip2pt\centerline{The Compact K\"ahler Complex Surfaces}}}
\dooley
\vskip .5in
\submit
The structure of topological quantum field theories on the
compact K\"ahler manifold is interpreted. The BRST transformations of fields
are derived from universal bundle and the observables are found from the
second Chern class of universal bundle.
\Date{}

\newsec{Introduction}
Witten introduced the relativistic quantum field theories with the
supersymmetry transformation called topological quantum field theory(TFT)
in order to explain Donaldson polynomial invariance of 4-dimensional
manifolds\wit. This supersymmetry is simular to BRST symmetry. So TFT can be
constructed by BRST gauge fixing to the space of anti-self dual connections
of some classical theory with the trivial Lagrangian\lab.
This BRST transformation laws are simply derived from universal bundle
theory. The BRST generator is the exterior derivative in part of the orbit
space, $\CA/\CG$  of the base manifold for the universal bundle\ati\singer\ken.
One can derive BRST transformation laws of the connection fields and ghost
fields from the curvature form and Bianchi identity of the universal
bundle. With another way twisting a set of $N=2$ spinorial supercharges
one could obtain the topological BRST generator\bir\witt.
Atiyah and Jeffrey showed that Witten action of Donaldson theory can
be interpreted by Mathai-Quillen formula about Euler number of
an infinite dimensional vector bundle of self-dual two forms over the
orbit space\aj.

Until now TFT was constructed on Riemann manifolds. However, we want to
extend TFT to the compact K\"ahler manifolds.
The moduli spaces, according to the bundle structure over
K\"ahler manifolds, have a geometric structure.
Itoh \itoi\itoii\ introduced K\"ahler structure in the moduli spaces of
anti-self dual connections in $SU(n)$-bundles over the compact K\"ahler
surfaces and Kim \kim\ introduced complex structures in the moduli spaces of
Einstein-Hermition vector bundles over compact K\"ahler manifolds.

In this letter, we show that TFT on the compact K\"ahler manifolds
can be constructed. In sec. 2 we construct BRST symmetry and
give Witten action over the moduli space of anti-self dual connections
and over the moduli space of Einstein-Hermitian connections.
In sec. 3 we derive the observables and show that the observables
are bigraded into holomorpic and anti-holomorpic parts.
Lastly, we have simple comments.

\newsec{The Action of TFT in BRST Formalisim}
On an $n$-dimensional complex manifold $M$, the exterior derivative can be
decomposed into the holomorphic and the anti-holomorphic part,
$d=\rd +\bar\rd$. Also, any exterior $r$-form can be decomposed into
$(p,q)$-forms such that $p+q=r$.

Let $P\to M$ be a principal $G$-bundle over $M$ and $\CG$ be the group
of gauge transformations of $P$. Locally faithful representation $\r$ of $G$
to $GL(r,C)$ induces a smooth complex vector bundle $E=P\times_{\r}C^r$ over
$M$. One calls $E$ a smooth complex vector bundle of rank $r$ over $M$.
We will use the following;\hfil\break
$A^{p,q}=$ the space of $(p,q)$-forms over $M$\hfil\break
$A^{p,q}(E)=$ the space of $(p,q)$-forms over $M$ with values in $E$\hfil\break
so that
\eqn\lei{
A^r=\sum_{p+q=r}A^{p,q},\qquad A^r(E)=\sum_{p+q=r}A^{p,q}(E).
}
We define an inner product in the space of $(p,q)$-form on $M$ such that
\eqn\leew{
(\f,\p)=\int_M \f\w\bar\ast\p,\qquad for\ given\ (p,q)-forms\ \f\ and\ \p,
}
where $\bar\ast$ is the dual operator defining
\eqn\leqw{
\bar\ast:A^{p,q}\to A^{n-p,n-q}.
}

The covariant derivative $D$ in $E$ may be decomposed into $D=\rd_A+\bar\rd_A$
such that $\rd_A:A^{p,q}(E)\to A^{p+1,q}(E)$ and
$\bar\rd_A:A^{p,q}(E)\to A^{p,q+1}(E)$.
We can decompose the curvature $2$-form  which is well
known in a K\"ahler manifold,
\eqn\leii{
B^2=B^2_+ \oplus B^2_-
}
where $B^2_+ =\{F^{2,0}+F^{0,2}+f\o|f\equiv\L F^{1,1}\ and\ \o=K\"ahler\
form\}$
and $B^2_-$ consists of $\CG$-valued  real $(1,1)$-forms .
Here $\L$ is the adjoint, $(\f,\L\p)=(L\f,\p)$ of operator $L$ which denotes
the exterior multiplication by K\"ahler form $\o$ so that
\eqn\leiii{
L:A^{p,q}\to A^{p+1,q+1}.
}

We consider the set $\CD$ of all possible connections on the principal
$G$-bundle, $P$, over a base manifold $M$ and the set $\CG$ of the complexified
group of gauge transformations of $P$. $\CD/\CG$ denotes the gauge equivalence
classes of connections called orbit space.

We define the universal bundle $\CQ$ over the base manifold $M\times\CD/\CG$.
$\CQ$ is the total space of a principal $G$-bundle,
\eqn\uuu{
\CQ\to \CQ/G=M\times\CD/\CG
}
over $M\times\CD/\CG$.

We denote $A^{p,q,r}(M\times\CD/\CG)$ the space of $(p,q)$-forms on the complex
manifold $M$ and $r$-forms on $\CD/\CG$.
On the base manifold $M\times\CD/\CG$, the exterior derivative can be
decomposed into $\bar d=\rd+\bar\rd+s$ such that
\eqn\leiv{\eqalign{
& \rd:A^{p,q,r}(M\times\CD/\CG)\to A^{p+1,q,r}(M\times\CD/\CG)\cr
& \bar\rd:A^{p,q,r}(M\times\CD/\CG)\to A^{p,q+1,r}(M\times\CD/\CG)\cr
& s:A^{p,q,r}(M\times\CD/\CG)\to A^{p,q,r+1}(M\times\CD/\CG),\cr}
}
if the manifold $M$ keeps the complex manifold.
The connections of the universal bundle may be decomposed into
$\bar A=A'+A''+C$  that is $A'\in A^{1,0,0},\ A''\in A^{0,1,0}$ and
$C\in A^{0,0,1}$. We identify the curvature components with the curvature
multiplet $(F^{2,0},F^{0,2},F^{1,1},\p,\bar\p,\f)$ satisfying the horizontal
condition. The ghost fields, $\p,\bar\p$ and $\f$, describe differential forms
on the orbit space.

we can obtain the following transformation laws from the curvature formula
and Bianchi identity for the universal bundle over $M\times\CD/\CG$ ;
\eqn\li{\eqalign{
sA' &=\p,\cr
sA'' &=\bar\p,\cr
s\p &=-\rd_A \f,\cr
s\bar\p &=-\bar\rd_A\f,\cr
s\f &=0\cr
}}
where $\rd_A=\rd +A'$ and $\bar\rd_A=\bar\rd +A''$.

In order to obtain the BRST invariant action, we need gauge fixing
conditions and the slice condition which is orthogonal to
the variations in connection $A$ that can be obtained by a gauge
transformation.
We will choose the anti-self dual conditions,
\eqn\tty{
F^{2,0}=F^{0,2}=\L F^{1,1}=0,
}
which are identical to the equations defining the moduli space of instantons,
$\CM$
and the slice condition,
\eqn\ttyi{
\L (\bar\rd_A\p -\rd_A\bar\p)=0.
}

We need additional multiplets to write Lagrangian. So we introduce two
pairs $(\l,\eta)$ and $(\bar\c,\bar H)$ such that
\eqn\abc{\eqalign{
& s\l=\eta,\qquad\quad s\eta=[\f,\l],\cr
& s\bar\c=\bar H, \qquad\quad s\bar H=[\f,\bar\c].\cr}
}
The multiplets $(\bar\c,\bar H)$ are self-dual 2-forms and have the ghost
number $(-1,0)$. So the self-dual 2-forms $\c$ and $H$ can be written as
\eqn\abd{\eqalign{
& \bar\c=\c^{2,0}+\c^{0,2}+\c\o,\cr
& \bar H=H^{2,0}+H^{0,2}+H\o,\cr}
}
where $\c=\L\bar\c$ and $H=\L\bar H$.
The multiplets $(\l,\eta)$ are zero forms and have the ghost number
$(-2,1)$.

We can obtain the action as
\eqn\liii{\eqalign{
\CS=& s\int_M Tr[\c^{0,2}\w\bar\ast (F^{0,2}+\Fr{1}{8}H^{0,2})
+\c^{2,0}\w\bar\ast(F^{0,2}+\Fr{1}{8}H^{2,0})+\c (\L F+\Fr{1}{4}H)
\o\w\bar\ast\o \cr
&+\l\L(\bar\rd_A \p -\rd_A \bar\p)\o\bar\ast\o]. \cr}
}

Then by using Eq.\li\  the action \liii\ can lead to
\eqn\lv{\eqalign{
\CS=&\int_M Tr[H^{0,2}\w\bar\ast (F^{0,2}+\Fr{1}{8} H^{0,2})
+H^{2,0}\w\bar\ast (F^{2,0}+\Fr{1}{8} H^{2,0})\cr
&+\c^{0,2}\w(\bar\ast\bar\rd_A\bar\p+\Fr{1}{8} [\f,\bar\ast\c^{0,2}])
+\c^{2,0}\w(\bar\ast\rd_A\p+\Fr{1}{8} [\f,\bar\ast\c^{2,0}])
+H(\L F+\Fr{1}{4} H)\o\w\bar\ast\o\cr
& +H\{ \L(\rd_A\bar\p +\bar\rd_A \p)+\Fr{1}{4} [\f,H]\} \o\w\bar\ast\o
+\{ \eta\L(\bar\rd_A\p-\rd_A\bar\p) \cr
& +\l(\bar\rd_A^*\rd_A +\rd_A^*\bar\rd_A)\f
+\L(\bar\p\w[\l,\p]-\p\w[\l,\bar\p])\}\o\w\bar\ast\o].\cr}
}

By eliminating an auxiliary fields $H$ the action \lv\ be rewritten as
\eqn\lv{\eqalign{
\CS=& \{ 2(||F^{2,0}||^2+||F^{0,2}||^2)+||\L F^{1,1}||^2 \}\cr
&+\int_M[\c^{0,2}\w(\bar\ast\bar\rd_A\bar\p+\Fr{1}{8} [\f,\bar\ast\c^{0,2}])
+\c^{2,0}\w(\bar\ast\rd_A\p+\Fr{1}{8} [\f,\bar\ast\c^{2,0}])
+H(f+\Fr{1}{4} H)\o\w\bar\ast\o\cr
& +H\{ \L(\rd_A\bar\p +\bar\rd_A \p)+\Fr{1}{4} [\f,H]\} \o\w\bar\ast\o
+\{ \eta\L(\bar\rd_A\p-\rd_A\bar\p) \cr
& +\l(\bar\rd_A^*\rd_A +\rd_A^*\bar\rd_A)\f
+\L(\bar\p\w[\l,\p]-\p\w[\l,\bar\p])\}\o\w\bar\ast\o].\cr}
}
Using the following identity\bra
\eqn\qqq{
|F|^2\o^{[n]} =|\L F|^2\o^{[n]}+Tr(F\w F)\w\o^{[n-2]}+2(|F^{0,2}|^2 +
|F^{2,0}|^2)\o^{[n]},
}
we can construct the action used by Witten, $Yang-Mills+\cdots$,
\eqn\lvv{\eqalign{
\CS=&\int_M [|F|^2\o^{[2]} -Tr(F\w F)] \cr
&+\int_M[\c^{0,2}\w(\bar\ast\bar\rd_A\bar\p+\Fr{1}{8} [\f,\bar\ast\c^{0,2}])
+\c^{2,0}\w(\bar\ast\rd_A\p+\Fr{1}{8} [\f,\bar\ast\c^{2,0}])
+H(f+\Fr{1}{4} H)\o\w\bar\ast\o\cr
& +H\{ \L(\rd_A\bar\p +\bar\rd_A \p)+\Fr{1}{4} [\f,\c]\} \o\w\bar\ast\o
+\{ \eta\L(\bar\rd_A\p-\rd_A\bar\p) \cr
& +\l(\bar\rd_A^*\rd_A +\rd_A^*\bar\rd_A)\f
+\L(\bar\p\w[\l,\p]-\p\w[\l,\bar\p])\}\o\w\bar\ast\o].\cr}
}
where $\int_M Tr(F\w F)$ is the second Chern character of $E$
and $\o^{[n]}=\Fr{1}{n!}\o^n$.
This action is invariant under the transformations;
\eqn\lli{\eqalign{
sA' =\p,\qquad\quad & sA'' =\bar\p,\cr
s\p =-\rd_A \f,\qquad\quad & s\bar\p =-\bar\rd_A\f,\cr
s\f =0,\qquad\quad  & s\c=2f,\cr
s\c^{2,0}=4F^{2,0},\qquad\quad & s\c^{0,2}=4F^{0,2},\cr
s\l =\eta,\qquad\quad & s\eta=[\f,\l].\cr }
}

Further we consider unitary connections on a complex bundle such that
$|F^{2,0}|^2=|F^{0,2}|^2$.
If the gauge fixing conditions is restricted on the connection space
satisfied Einstein-Hermitian equations,
\eqn\hhh{\eqalign{
&F^{0,2}=0\cr
&i\L F+cI=0,\cr}
}
where $c$ is constant number and $I\in\O^0(X,End\ E)$ is identity
section,  a similar constructing of topological field theories is
possible. In this sector the topological invariant action can be
constructed by
\eqn\lll{\eqalign{
\CS=& s\int_M Tr[\c^{0,2}\w\bar\ast (F^{0,2}+\Fr{1}{16}H^{0,2})
+\c (i\L F+cI+\Fr{1}{4}H) \o\w\bar\ast\o \cr
&+\l\L(\bar\rd_A \p -\rd_A \bar\p)\o\bar\ast\o]. \cr}
}

By using transformation laws \li\ and \abc, identity \qqq\ and
\eqn\sss{\eqalign{
||i\L F +cI||^2 &=||i\L F||^2+2<i\L F,I>+|c|^2<I,I>\cr
<i\L F,I> &={i\over {2\pi}}\int_X Tr(F,\o)\o^n=i\int_X TrF\w\o^{n-1},\cr}
}
and eliminating auxiliary fields $H$, one can find that the action \lll\
reachs by simple calculation
\eqn\llvv{\eqalign{
\CS=&\int_M [|F|^2\o^{[2]} -Tr(F\w F)-i TrF\w\o+|c|^2 I] \cr
&+\int_M[\c^{0,2}\w(\bar\ast\bar\rd_A\bar\p+\Fr{1}{16} [\f,\bar\ast\c^{0,2}])
+H(i\L F+cI+\Fr{1}{4} H)\o\w\bar\ast\o\cr
& +H\{ \L(\rd_A\bar\p +\bar\rd_A \p)+\Fr{1}{4} [\f,\c]\} \o\w\bar\ast\o
+\{ \eta\L(\bar\rd_A\p-\rd_A\bar\p) \cr
& +\l(\bar\rd_A^*\rd_A +\rd_A^*\bar\rd_A)\f
+\L(\bar\p\w[\l,\p]-\p\w[\l,\bar\p])\}\o\w\bar\ast\o].\cr}
}
where $\int_M TrF\w\o^{[n-2]}$ is the first Chern class of $E$.
This action is invariant under the transformations;
\eqn\lvi{\eqalign{
sA' =\p,\qquad\quad & sA'' =\bar\p,\cr
s\p =-\rd_A \f,\qquad\quad & s\bar\p =-\bar\rd_A\f,\cr
s\c=2(i\L F+cI), \qquad\quad & s\c^{0,2}=4F^{0,2},\cr
s\l =\eta,\qquad\quad & s\eta=[\f,\l],\cr
s\f=0.\cr}
}

\newsec{Observables}

In this section, we hope to find the topological invariant observables.
The observables must be BRST invariant, not depend explicitly on the
metric, and not be written as $s$-exact, $s\r$.
According to Galperin and Ogievetsky\gal\ these operators can be constructed
as certain operators
\eqn\leei{
I(\CO)=\int_M \CW_{2-p,2-q}\w\CO,
}
where $\CW_{p,q}$, $(p,q=0,1,2)$, is a $(2-p,2-q)$-form constructed
out of the fields and $\CO$ is an $(p,q)$-form on $M$ which does not
depend on fields and is closed, $d\CO=0$.
These are
\eqn\leeii{\eqalign{
& \CW^4_{0,0}=Tr\f^2,\cr
& \CW^3_{0,1}=Tr(2\bar\p\f),\cr
& \CW^2_{0,2}=Tr(\bar\p\w\bar\p+2F^{0,2}\f),\cr
& \CW^2_{1,1}=Tr(2\bar\p\w\p+2F^{1,1}\f),\cr
& \CW^3_{1,0}=Tr(2\p\f),\cr
& \CW^2_{2,0}=Tr(\p\w\p+2F^{2,0}\f),\cr
& \CW^1_{2,1}=Tr(2F^{2,0}\w\bar\p+2F^{1,1}\w\p),\cr
& \CW^1_{1,2}=Tr(2F^{0,2}\w\p+2F^{1,1}\w\bar\p),\cr
& \CW^0_{2,2}=Tr(2F^{2,0}\w F^{0,2}+F^{1,1}\w F^{1,1}).\cr}
}
These arise from the components $c^{(2,2)-i}$ of the second Chern class,
$c_2=Tr(\CF\CF)$, in the universal bundle,

We can easely show that
\eqn\leeiv{
\rd \CW^4_{0,0}=-s\CW^3_{1,0}, \qquad \bar\rd\CW^4_{0,0}=-s\CW^3_{0,1},
}
and one finds recursively
\eqn\leev{\eqalign{
\rd\CW^{4-p-q}_{p,q}=-s\CW^{4-p-1-q}_{p+1,q},& \qquad
\bar\rd\CW^{4-p-q}_{p,q}=-s\CW^{4-p-q-1}_{p,q+1}\cr
\rd\CW^0_{2,2}=0,  & \qquad \bar\rd\CW^0_{2,2}=0.\cr}
}
Then the integral $I(\CO)$ is BRST invariant, since
\eqn\rrr{
sI(\CO)=\int_M s\CW_{2-p,2-q}\w\CO=\int_{\g}\rd\CW_{2-(p-1),2-q}\w\CO=0
}
or
\eqn\rst{
sI(\CO')=\int_M s\CW_{2-p,2-q}\w\CO'=\int_M \bar\rd\CW_{2-p,2-(q-1)}\w\CO'=0.
}
Assume that $\CO=\rd\s$ or $\CO=\bar\rd\s$,
\eqn\jjj{\eqalign{
I(\CO) &=\int_M \CW_{2-p,2-q}\w\rd\s=(-1)^{p+q}\int_M
\rd\CW_{2-p,2-q}\w\s\cr
&=(-1)^{p+q}\int_M s\CW_{2-(p+1),2-q}\w\s=s((-1)^{p+q}\int_M
\CW_{2-(p+1),q}\w\s)\cr}
}
or
\eqn\jji{\eqalign{
I(\CO) &=\int_M \CW_{2-p,2-q}\w\bar\rd\s=(-1)^{p+q}\int_M
\bar\rd\CW_{2-p,2-q}\w\s\cr
&=(-1)^{p+q}\int_M s\CW_{2-p,2-(q+1)}\w\s=s((-1)^{p+q}\int_M
\CW_{2-p,2-(q+1)}\w\s).\cr}
}
The expectation values of such observables vanish
Therefore we can obtain the observables of topological BRST symmetry from
the de Rham cohomology classes of M.

Let us define the map
\eqn\nnn{
\P_{\CO}:\sum_{p+q=r} H^{p,q}(M)\to H^r(\CM),
}
which takes the form
\eqn\fff{
\P_{\CO}=\int_{M}\CW_{2-p,2-q}\w\CO
}
where $\CO$ is a $\bar\rd$-closed or $\rd$-closed $(p,q)$-form on $M$.
We can see that such a map corresponds to Donaldson map defineing over
cohomology.

\newsec{Conclusion}
We constructed Witten action and topological invariant observables on
the compact K\"ahler manifolds. These results were realised from the
universal bundle structures. The BRST transformations of the fields were
obtained by the curvature 2-form and Bianchi identity. The second Chern class
in the universal bundle leads to the topological invariant observables.
We concerned only on the structure which  preserved the complex structure
of $M$ in the base manifold $M\times \CD/\CG$ of the universal bundle.

If the orbit space $\CD/\CG$ is restricted to the subspace $\CH/\CG$ which has
anti-self dual connections, $\CH/\CG$ has the K\"ahler structure. The exterior
derivative of $\CD/\CG$ which equals to the BRST generator may be decomposed
into the holomorpic and the anti-holomorpic parts, $s=s'+s''$.
Then we have two kind of BRST symmetry. The universal bundle with the K\"ahler
structure on both $M$ and $\CH/\CG$ has various structures. In the special
case, Park\park\ showed that the universal bundle gives the $N=2$ symmetry.
However, one needs further to study the universal structure over the base
manifold, $M\times\CH/\CG$.

\bigbreak\bigskip\bigskip\centerline{{\bf Acknowledgements}}
The author would like to express his gratitude to Professor J. H. Yee and
J. S. Park for helpful discussions. This work was
supported in part by the Center for Theoretical Physics (SNU), the Korea
Science and Engineering Foundation, the Ministry of Education and Daewoo
Foundation.

\listrefs
\end